\documentclass[12pt]{article}
\usepackage{graphicx}
\begin {document}
\title {MODIFIED LANDAUER PRINCIPLE ACCORDING TO TSALLIS ENTROPY}

\author{Luis Herrera \\Instituto Universitario de F\'isica
Fundamental y Matem\'aticas, \\Universidad de Salamanca, Salamanca 37007, \\Spain \thanks{ E-mail address: lherrera@usal.es}}
\date{}
\maketitle

\abstract{The Landauer principle establishes a lower bound in the amount of energy  that should be dissipated in the erasure of one bit of information. The specific value of this dissipated energy is tightly related to the definition of entropy. In this article, we present a generalization of the Landauer principle based on the Tsallis entropy. Some consequences resulting from such a generalization are discussed. 
{These consequences include the modification to the mass  ascribed to one bit of information,  the generalization  of the Landauer principle to the case when the system is embedded  in a gravitational field}, and the number of bits radiated in the emission of gravitational waves.}

\section{Introduction}

The Landauer  principle Ref. \cite{Lan}, which is a cornerstone in the theory of information, states that the erasure of  one bit of information stored in a system implies  the dissipation   of energy, whose value cannot be   smaller than 

\begin{equation}
\bigtriangleup E=kT \ln2,
\label{lan1}
\end{equation}
where $k$ is the Boltzmann constant and $T$ denotes the  temperature of the environment. {The important point to keep in mind here is that even though the value of dissipated energy depends on the erasure procedure,   it cannot be lower that Equation (\ref{lan1})}.

At this point, some remarks are needed, as follows:

\begin{itemize}
\item It has been argued in the past  (see for example Ref. \cite{7}) ) that the main idea stated in the Landauer principle appears already in some Brillouin works Ref. \cite{3}. We shall skip this controversy  {and shall adopt  the notation used by most of researchers, and we shall refer  to it as the Landauer principle}.
\item {In spite of some arguments put forward in the past questioning the relevance  of the Landauer  principle  (see Ref. \cite{B} and references therein), the important point to retain here is that on  the one hand   it allows an ``informational'' reformulation of thermodynamics, as stressed in Ref. \cite{B}, and on the other hand  brings out a  link between information theory and different branches of  science Refs. \cite{Plenio,   bais, BII}. This allows us to approach    some  physical problems from the point of view of information theory.}
\item The expression Equation (\ref{lan1}) for the dissipated energy heavily relies on the concept of entropy. More specifically, such an expression was found using the Gibbs entropy. Therefore, we should expect different expressions for alternative definitions of entropy.
\end{itemize}

\section{Landauer Principle and Definition of Entropy}
In order to exhibit the link between the Landauer principle and the definition of entropy, let us present a very simple  proof of this principle.

Thus, let us consider a physical system which may be in two possible states, e.g., a particle whose spin may point upward or downward. The particle is inside a black box, and for an observer outside the box, the particle may be in either state with the same probability. Then, using the Gibbs definition of entropy given by
\begin{equation}
S=-k\sum^{N}_{i=1}{p_i\ln p_i},
\label{2n}
\end{equation}
where $N$ denotes the total number of accessible states and $p_i$ is the probability of each state (i.e.,  $\sum^{N}_{i=1}{p_i}=1$), we find that the Gibbs entropy of  our system is 
\begin{equation}
S=k\ln{2}.
\label{3}
\end{equation}

Let us now apply a magnetic field to our system as a consequence of which the spin is set to point to a determined direction (upward or downward). Obviously after such operation the entropy of the system becomes equal to zero, implying that there has been a decreasing of entropy equal to
\begin{equation}
\Delta S=k\ln{2},
\label{4}
\end{equation}
which according to the second law of thermodynamics should be accompanied by an increasing of,  at least,  the same amount, producing a dissipation of  energy equals to
\begin{equation}
\Delta E=kT\ln{2},
\label{5}
\end{equation}
where $T$ is the temperature of the environment.

Now,   applying a magnetic field to our system, we set  the direction of the spin in  a predetermined direction, thereby   erasing the  information about where the spin was pointed to  before switching on the magnetic  field Ref. \cite{pie}.  Since this information is contained in the answer to the single question, "where is the spin  pointing to?",  
the amount of this information is one bit. 

Thus, we have proved that erasing one bit of information implies that an amount of energy not smaller than Equation (\ref{5}) must be dissipated, which is just the statement of the Landauer principle. The purpose of the above exercise being to bring out the relationship between the minimal amount of  dissipated energy with the definition of entropy \mbox{Equation (\ref{2n}).}

Arriving at this point the obvious question arises:  what could be the corresponding minimal amount of dissipated energy in the process of erasure of one bit of information if we resort to a definition of entropy different from Equation (\ref{2n})? 

{We endeavor in  this work to answer to the above question  in the case when  we use the Tsallis entropy (instead of using Equation (\ref{2n})).} 

{However, it would be legitimate to ask why, in particular, we have chosen Tsallis entropy, instead of any other  definition of entropy?  The answer to this question is based on the great deal of attention received by Tsallis proposal and its applications (see for example Ref. \cite{tsr1,tsr5,tsr2,tsr3,tsr4} and references therein). Nevertheless, it goes without saying, that resorting to any other alternative definition of entropy would also  deserve to be considered}.

\section{Tsallis Entropy and Modified Landauer Principle}
Some years ago Tsallis proposed a generalization of Gibbs definition of entropy, which reads Ref. \cite{Ts}
\begin{equation}
S=k\frac {1-\sum^{N}_{i=1}{p^q_i}}{q-1},
\label{6n}
\end{equation}
where $q$ is a real number. 

It is a simple matter to check that 
\begin{equation}
\lim_{q\rightarrow 1}k\frac {1-\sum^{N}_{i=1}{p^q_i}}{q-1}=-k\sum^{N}_{i=1}{p_i\ln p_i}.
\label{7n}
\end{equation}

Thus, deviations from the Gibbs entropy correspond to values of $q$ different from $1$.

Since its publication the Tsallis proposal has received a great deal of attention, and therefore we find it useful to evaluate its impact in the Landauer principle.
\subsection*{The Lower Bound of the Dissipated Energy Ensuing  the Erasure of One Bit of Information According to the Tsallis Entropy}
In order to calculate the minimal amount of energy that must be dissipated when erasing on bit of information  according to Tsallis entropy, let us retrace the steps of the exercise proposed in the previous section, leading to Equations (\ref{4}) and (\ref{5}). 

\begin{figure}
\includegraphics[width=10cm]{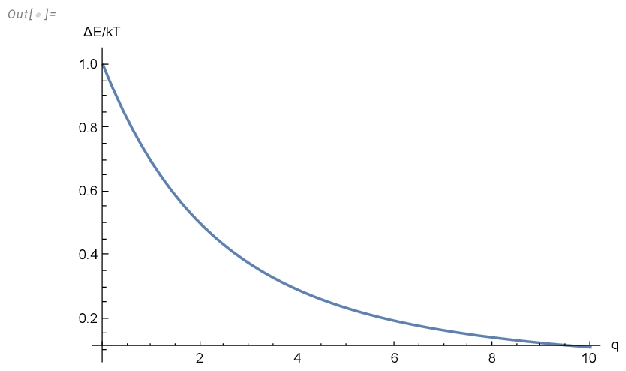}
\caption {$\Delta E/kT$ as function of $q$ for the Tsallis entropy.}\label{fig1}
\end{figure}

{Thus, let us  consider a system with two possible accessible states ($N=2$) the probability of each of which is $1/2$. Then, it follows from Equation (\ref{6n}) that the Tsallis entropy of our system is given by
 \begin{equation}
 S=\frac{k}{q-1}\left[1-2^{(1-q)}\right].
\label{7nn}
\end{equation}
We now proceed to apply a  magnetic field to our system, after which the system is in a  single state with probability $1$, implying the erasure of one bit of information, and the vanishing of the entropy.  Thus,  the decreasing of entropy is given  by 
 \begin{equation}
\Delta S=\frac{k}{q-1}\left[1-2^{(1-q)}\right],
\label{8n}
\end{equation}
producing an amount of dissipated  energy equal to
 \begin{equation}
\Delta E\equiv T\Delta S=\frac{kT}{q-1}\left[1-2^{(1-q)}\right].
\label{9n}
\end{equation}
As depicted in Figure \ref{fig1}, the above expression decreases  monotonically with $q$ for any $q>0$.}
It is a simple matter to check that in the limit $q\rightarrow 1$, expressions Equations (\ref{8n}) and (\ref{9n}) become Equations (\ref{4}) and  (\ref{5}), respectively.

Thus, using the Tsallis entropy we see that the minimal energy dissipated in the erasure of one bit of information depends on the parameter $q$ as expressed by (\ref{9n}). 

We shall next see how this change affects some consequences derived from the Landauer principle.

\section{The Mass of a Bit of Information}
As we have seen above, the Landauer principle  based on the Gibbs definition of entropy,  asserts that a minimal amount of energy given by Equation (\ref{lan1}),  should be dissipated when erasing one bit of information. {This fact implies the association of such an amount of energy with one bit of information. From the previous  comment it follows at once, recalling  the well-known fact  that a mass $E/c^2$ has to be ascribed  to energy $E$, and that a mass should be  ascribed to a bit of information  Ref. \cite{3m} (see also Refs. \cite{B,vop}), specifically}
\begin{equation}
\bigtriangleup M=\frac{kT }{c^2}\ln2,
\label{lan11}
\end{equation}
where $c$ denotes the speed of light.

Following  the same reasoning leading to Equation (\ref{lan11}), but using Equation (\ref{9n}) instead of Equation (\ref{lan1}) we obtain for the mass associated with a bit of in formation
 \begin{equation}
\Delta M=\frac{kT}{c^2(q-1)}\left[1-2^{(1-q)}\right],
\label{9nn}
\end{equation}
which of course becomes Equation (\ref{lan11}) in the limit $q\rightarrow 1$.

From the above it follows that  for one bit of information, at room temperature, the minimal dissipated energy is 
\begin{equation}
\bigtriangleup E \approx \frac{4.04 }{q-1}\left[1-2^{(1-q)}\right] \times 10^{-14} erg
\label{lan1bis}
\end{equation}
and  the associated mass is:
\begin{equation}
\bigtriangleup  M\approx  \frac{4.33}{(q-1)}\left[1-2^{(1-q)}\right]\times 10^{-35} g.
\label{lan111}
\end{equation}

In the limit $q\rightarrow 1$, the above expressions yield $2.8\times 10^{-14} erg$ and $3\times 10^{-35} g.$, respectively.

{Also it is worth noticing that according to  the uncertainty principle, there is a minimal time interval required to measure  a given  amount of energy. In our case this implies that for the energy  Equation (\ref{lan1bis}) the minimal time interval is}
\begin{equation}
\bigtriangleup t\approx\frac{\hbar}{\bigtriangleup E}\approx \frac{2.56 (q-1)}{1-2^{1-q}} \times 10^{-14} s ,
\label{la7bis}
\end{equation}
where $\hbar$ is the Planck constant divided by $2\pi$,
thereby imposing a limit in the speed of information processing which in the case $q\approx 1$ is $\approx 10^5GHz$.

We shall next consider the case, when the system is placed in a gravitational field.
\section{Landauer Principle in  a Gravitational Field}

If the system is located in a (weak) static gravitational field, then the gravitational potential affects the Landauer principle. This important result was obtained by  Daffertshoffer and  Plastino Ref. \cite{pla}. More specifically, these authors show that in this case (assuming for the entropy the Gibbs definition Equation (\ref{2n})) the minimal amount of energy dissipated in the erasure of one bit of information is given by
\begin{equation}
\bigtriangleup E=kT(1+\frac{\phi}{c^2}) \ln2. 
\label{lan2}
\end{equation}
where $\phi$ denotes the (negative) gravitational potential, and  $T(1+\frac{\phi}{c^2})$ (the Tolman's temperature) is the quantity  which is constant  in thermodynamic equilibrium Ref. \cite{Tol}.

Now, Equation (\ref{lan2}) was obtained in the context of Newtonian gravity (weak field approximation). {The extension of the above result to  the general relativistic case is simple to achieve if we recall that  in such a case   Tolman's temperature becomes $T\sqrt{g_{tt}}$, where $g_{tt}$ denotes the $tt$ component of the metric tensor (the coefficient of $dt^2$ in the expression for the line element)}. Therefore,  Equation (\ref{lan2})  generalizes to 
\begin{equation}
\bigtriangleup E=kT \sqrt{g_{tt}}\ln2,
\label{lan3}
\end{equation}
producing for the mass ascribed to a bit of information 
\begin{equation}
\bigtriangleup M =\frac{kT}{c^2} \sqrt{g_{tt}}\ln2.
\label{lan4}
\end{equation}

So, the question is: how does the Landauer principle change in the presence of a gravitational field if we use the Tsallis entropy Equation (\ref{6n}) instead of Equation (\ref{2n})?

Retracing the same steps followed in Ref. \cite{pla}, we obtain  for the energy dissipated in the erasure of one bit of information and for the corresponding mass
 \begin{equation}
\Delta E=\frac{kT}{q-1}\left[1-2^{(1-q)}\right](1+\frac{\phi}{c^2}),
\label{9nt}
\end{equation}

and 
 \begin{equation}
\Delta M=\frac{kT}{c^2(q-1)}\left[1-2^{(1-q)}\right](1+\frac{\phi}{c^2}).
\label{9nnt}
\end{equation}

{The generalization of the above expressions to the general relativistic case  may be easily obtained by replacing $(1+\frac{\phi}{c^2})$ with $\sqrt{g_{tt}}$.} 

{After the formation of a black hole   ($g_{tt}=0$), it follows from Equation (\ref{lan3}) or \mbox{Equation (\ref{9nt}}) that  the energy dissipated during the erasure of one bit of information   vanishes }(assuming that the proper temperature is not singular),  leading to a vanishing mass for a bit of information.

{Now, the change of information  without dissipation implies that   all  bits are already in one state only Ref. \cite{pie}. This result agrees with the well-known  assumption that the quantum radiation  emitted by the black hole   is nearly thermal (i.e., it conveys no information) Refs. \cite{H1, H2}, thereby suggesting the ``bleaching'' of information at the horizon. }

{Thus, the well-known fact that  after the formation of the horizon  ($g_{tt}=0$), no further information  leaves the system,  follows in a simple way from  information theory. }

{If the gravitational field does not correspond to a black hole ($g_{tt}\neq 0$), then we see a decreasing of  the corresponding mass of a bit of information.  Such a decreasing value depending on the parameter  $q$ occurs if we assume the Tsallis definition of entropy. 
At any rate such a decreasing value is very small for a weak gravitational field  ($\frac{\phi}{c^2}\approx 10^{-9}$ for the case of the earth).}

\section{Gravitational Radiation, Radiated Information, and the Landauer Principle}
Finally, we would like to consider   the relationship between the energy  and the information conveyed by gravitational radiation and the definition of entropy. 

{As we know from field theory (at classical level and for any spin), information about changes in the structure and/or  state of motion  of the source is propagated by radiation.} Once the   observers have received this information, the information encrypted in   the ``old'' multipole structure  is erased. In other words, the process of radiation implies not only  propagation of information but also  erasure of information, from which it is obvious  that the Landauer principle should be implicated in the whole process. 

 The above comments imply that   according to the Landauer principle, gravitational radiation  entails a dissipation of energy. This  conclusion  was proved to be true in Ref. \cite{out} and is a consequence of the fact that   gravitational radiation   is an irreversible process, and this irreversibility   should show up in the equation of state of the source. 

This ``informational'' approach to radiation is particularly manifest   in   the Bondi formalism Refs. \cite{bondi, 17}. 

{In this  approach there is a function (called ``news function'' by Bondi), 
which entails all  the information required  to forecast the evolution of the system (besides the initial data) and is  identified with gravitational radiation itself. Such an identification is possible because the news function describe all changes in the field  produced by changes in the source. Moreover, the vanishing of   the news function is the necessary and sufficient condition for the  total energy of the system to be constant.}  This scheme applies to Maxwell systems in Minkowski spacetime Ref. \cite{janis} as well as to Einstein--Maxwell systems Ref. \cite{18}.

{Once we admit that   a bit of radiated information  implies a bit of erased information at the radiating system, which in turn  leads to  a decreasing  of its total mass (energy), then it is legitimate to ask:  what part of the total radiated energy (mass)  corresponds to the radiated~information? }

We shall answer to this question, adopting   the Tsallis definition of entropy.

 In Ref. \cite{en} an answer was provided to the above question, based in the Landauer principle expressed through the Gibbs entropy Equation (\ref{lan1}). 

Thus, one obtains for the total dissipated energy (see Ref. \cite{en} for details)
\begin{equation}
E^{(L)}_{rad}=\int^\infty_{r_\Sigma}\int^\pi_0\int^{2\pi}_0{\sqrt{\vert g\vert}\mu^{(L)}_{rad}drd\theta d\phi},
\label{rad111}
\end{equation}
where $\vert g\vert$ is the absolute value of the determinant of the metric tensor,  $r=r_\Sigma$ is the equation of the boundary surface of the source, and $\mu^{(L)}$ is  the energy--density of the radiation associated  exclusively  with the dissipative processes related to the emission of gravitational~radiation.

The above expression may be transformed further using a central result by Bondi  Ref.~\cite{bondi}, relating the rate at which the energy is being radiated, with the news function,  which  reads:
\begin{equation}
\frac{dm(u)}{du}=-\frac{1}{2}\int^\pi_0{\frac{(dc(u, \theta))^2}{du}\sin \theta d\theta},
\label{new1}
\end{equation}
where $\frac{dc(u, \theta)}{du}$ is the news function, $u$ is the timelike coordinate in the Bondi frame,  $c(u,\theta)$ is a function entering into the power series expressions of the Bondi metric, and $m(u)$ denotes the energy of the system (the Bondi mass).

Therefore, the total radiated energy in the timelike interval $u_1\leq u \leq u_2$ is given by

\begin{equation}
E^{(L)}_{rad}=\int^{u_2}_{u_1}\left[  \frac{1}{2}\int^\pi_0{\frac{(dc(u, \theta))^2}{du}\sin \theta d\theta}\right]du,
\label{new2}
\end{equation}
(please notice a misprint in the sign of Equations (31) and (32) in Ref. \cite{en}).

On the other hand, according to the Landauer principle  Equation (\ref{9nt}),  we obtain for the total number $N$ of bits erased (radiated) in the process of the emission of gravitational~radiation

\begin{equation}
N= \frac{E^{(L)}_{rad}}{kT\sqrt{\vert g_{tt}\vert} \ln2}, 
\label{rad1111}
\end{equation}

Feeding back Equation 
(\ref{new2}) into Equation (\ref{rad1111}) we find an explicit relationship linking  the news function with the total number of bits radiated in the assumed time interval,
\begin{equation}
N= \frac{\int^{u_2}_{u_1}\left[  \frac{1}{2}\int^\pi_0{\frac{(dc(u, \theta))^2}{du}\sin \theta d\theta}\right]du.}{kT\sqrt{\vert g_{tt}\vert }\ln2}=\frac{\left[ \int^\infty_{r_\Sigma}\int^\pi_0\int^{2\pi}_0{\sqrt{\vert g\vert}\mu^{(L)}_{rad}drd\theta d\phi}\right]}{kT\sqrt{\vert g_{tt}\vert} \ln2},
\label{new3}
\end{equation}
which measure the total erased information during the radiation process.

The expressions above have been obtained  by resorting  to the Landauer  principle based on the Gibbs entropy; therefore, in the context of this work it is legitimate to ask how do the expressions above change if we use the Landauer principle based in the Tsallis entropy Equation (\ref{6n}). Using Equation (\ref{9nt}) and retracing the same steps leading to  Equation~(\ref{new3}), we obtain

\begin{equation}
N= \frac{(q-1)\int^{u_2}_{u_1}\left[  \frac{1}{2}\int^\pi_0{\frac{(dc(u, \theta))^2}{du}\sin \theta d\theta}\right]du.}{kT\sqrt{\vert g_{tt}\vert }\left[1-2^{(1-q)}\right]}=\frac{\left[ \int^\infty_{r_\Sigma}\int^\pi_0\int^{2\pi}_0{\sqrt{\vert g\vert}\mu^{(L)}_{rad}drd\theta d\phi}\right](q-1)}{kT\sqrt{\vert g_{tt}\vert} \left[1-2^{(1-q)}\right]},
\label{new3n}
\end{equation}
bringing out the role played by the parameter $q$ in the number of bits radiated in a given burst of gravitational radiation.

{It would be most desirable  to relate the above expressions with the data obtained from the LISA program (see Ref. \cite{lisa} and references therein). Unfortunately, at this point we do not see how to exactly establish such a link.}

\section{Discussion}
Motivated by the fact that the specific value of the lower bound of energy---which according Landauer principle should be dissipated in the erasure of one bit of information---depends on the definition of entropy, we have addressed the question about the value of this bound for the Tsallis entropy, obtaining the expression Equation (\ref{9n}).

Once this value has been  established, we have considered how deviations of this value, with respect to  the one obtained from the Gibbs entropy, affects  different scenarios where the Landauer principle is involved. {In particular we have brought out how different values of $q$ modify the values of different observational variables.}

{The first important result resides in the expression for the lower bound of energy dissipated after the erasure of one bit of information for the Tsallis entropy, which is now given by  Equation (\ref{9n}). Figure \ref{fig1} shows that for any value of temperature, such dissipated energy is a monotonically decreasing function of $q$, which is larger than the corresponding value for the Gibbs entropy for any value of $q$  in the interval $[0,1]$ and smaller in the interval $[1,\infty]$}.

Next, we have  considered  the mass associated with a bit of information, which for the Tsallis entropy is given by Equation (\ref{lan111}) in contrast with expression Equation (\ref{lan11}) obtained from the Gibbs entropy. This result also affects the limitation on the speed of processing, as expressed by Equation (\ref{la7bis}).

The generalization of the Landauer principle for systems embedded in a gravitational field has been achieved following the work by Daffertshoffer and  Plastino Ref. \cite{pla}. The corresponding expression for the energy dissipated in the erasure of one bit of information is now given by Equation (\ref{9nt}), leading to the expression Equation (\ref{9nnt}). {Once again we see how $q$ affects the values of the two above mentioned variables.}

Finally, we addressed the question about the number of bits radiated (erased) in the emission of gravitational radiation.  By using the Tsallis entropy, we found that such a number is given by Equation (\ref{new3n}) instead of the expression Equation (\ref{new3}) corresponding to the Gibbs entropy.

{In all these examples  the role of the parameter $q$ is clearly displayed}. This fact brings us back to the leitmotiv  of our work.

 Indeed, it  is to be expected that for any physical scenario, the experimental data could differentiate between what is the correct definition of entropy that should be adopted. In the case of Tsallis entropy, this implies a specific value of $q$. Since the  scenarios analyzed above imply observed quantities, we harbor the hope that some of the expressions  found here could help in a process of verification of the appropriate definition of entropy. {Moreover, we believe that the extension of the program followed in this work to other definitions of entropy is an issue that deserves to be considered in the future.}

\end{document}